\documentclass[
 reprint,
 bibnotes,
 amsmath,amssymb,
 aps,
 prl,
 ]{revtex4-2}
\usepackage{graphicx}
\usepackage{dcolumn}
\usepackage{bm}
\usepackage[pdfstartview=FitH,
CJKbookmarks=true,
bookmarksnumbered=true,
bookmarksopen=true,
colorlinks,
linkcolor=blue,
anchorcolor=blue,
citecolor=blue,
allcolors=blue,
pdfborder=001,
]{hyperref}
\usepackage[caption=false]{subfig}
\usepackage{floatrow}
\usepackage{enumitem}
\DeclareSubrefFormat{myparens}{#1~(#2)}
\DeclareCaptionListOfFormat{myparens}{#1(#2)}
\usepackage{xspace}
\usepackage{multirow}

\usepackage{cleveref}    
\crefrangeformat{figure}{#3#1#4--#5#2#6}
\Crefrangeformat{figure}{#3#1#4--#5#2#6}
\crefname{figure}{Fig.}{Figs.}
\Crefname{figure}{Fig.}{Figs.}
\usepackage{placeins}

\begin{document}

\newcommand\MeV{\ensuremath{\mathrm{MeV}}}
\newcommand\GeV{\ensuremath{\mathrm{GeV}}}
\newcommand*{\Mpi}{\ensuremath{1828 \pm 8 ({\rm stat})}\xspace}
\newcommand*{\Wpi}{\ensuremath{638 \pm 26 ({\rm stat})}\xspace}
\newcommand*{\Bpi}{\ensuremath{4.30 \pm 0.14 ({\rm stat})}\xspace}
\newcommand*{\MpisysR}{\ensuremath{^{+11}_{-33}}\xspace}
\newcommand*{\WpisysR}{\ensuremath{^{+35}_{-86}}\xspace}
\newcommand*{\BpisysR}{\ensuremath{^{+1.04}_{-1.03}}\xspace}
\newcommand*{\BpisysRP}{\ensuremath{^{+24}_{-24}}\xspace}
\newcommand*{\Mpisys}{\ensuremath{\MpisysR ({\rm syst})}\xspace}
\newcommand*{\Wpisys}{\ensuremath{\WpisysR ({\rm syst})}\xspace}
\newcommand*{\Bpisys}{\ensuremath{\BpisysR ({\rm syst})}\xspace}
\newcommand*{\MpiC}{\ensuremath{1675 \pm 9 ({\rm stat})}\xspace}
\newcommand*{\WpiC}{\ensuremath{443 \pm 15 ({\rm stat})}\xspace}
\newcommand*{\BpiC}{\ensuremath{4.27 \pm 0.14 ({\rm stat})}\xspace}
\newcommand*{\MpisysRC}{\ensuremath{^{+32}_{-32} }\xspace}
\newcommand*{\WpisysRC}{\ensuremath{^{+18}_{-56} }\xspace}
\newcommand*{\BpisysRC}{\ensuremath{^{+0.86}_{-1.32}}\xspace}
\newcommand*{\BpisysRPC}{\ensuremath{^{+20}_{-31}}\xspace}
\newcommand*{\MpisysC}{\ensuremath{\MpisysRC ({\rm syst})}\xspace}
\newcommand*{\WpisysC}{\ensuremath{\WpisysRC ({\rm syst})}\xspace}
\newcommand*{\BpisysC}{\ensuremath{\BpisysRC ({\rm syst})}\xspace}
\newcommand*{\Bpiwithchic}{\ensuremath{4.19 \pm 0.14 ({\rm stat})}\xspace}
\newcommand*{\BpisysRwithchic}{\ensuremath{^{+1.01}_{-1.00}}\xspace}
\newcommand*{\BpisysRPwithchic}{\ensuremath{^{+24.1}_{-23.8}}\xspace}
\newcommand*{\Bpisyswithchic}{\ensuremath{\BpisysRwithchic ({\rm syst})}\xspace}
\newcommand*{\Mpipole}{\ensuremath{1690 ^{+16}_{-16} ({\rm stat})}\xspace}
\newcommand*{\Wpipole}{\ensuremath{433 ^{+10}_{-11} ({\rm stat})}\xspace}
\newcommand*{\MpisysRpole}{\ensuremath{ ^{+36}_{-44}}\xspace}
\newcommand*{\WpisysRpole}{\ensuremath{^{+14}_{-39}}\xspace}
\newcommand*{\Mpisyspole}{\ensuremath{\MpisysRpole ({\rm syst})}\xspace}
\newcommand*{\Wpisyspole}{\ensuremath{\WpisysRpole ({\rm syst})}\xspace}
\newcommand*{\Mpipoleconst}{\ensuremath{1689 \pm 10 ({\rm stat})}\xspace}
\newcommand*{\Wpipoleconst}{\ensuremath{439 \pm 15 ({\rm stat})}\xspace}
\newcommand*{\MpisysRpoleconst}{\ensuremath{^{+33}_{-18}}\xspace}
\newcommand*{\WpisysRpoleconst}{\ensuremath{^{+35}_{-55}}\xspace}
\newcommand*{\Mpisyspoleconst}{\ensuremath{\MpisysRpoleconst ({\rm syst})}\xspace}
\newcommand*{\Wpisyspoleconst}{\ensuremath{\WpisysRpoleconst ({\rm syst})}\xspace}

\newcommand{\BESIIIorcid}[1]{\href{https://orcid.org/#1}{\hspace*{0.1em}\raisebox{-0.45ex}{\includegraphics[width=1em]{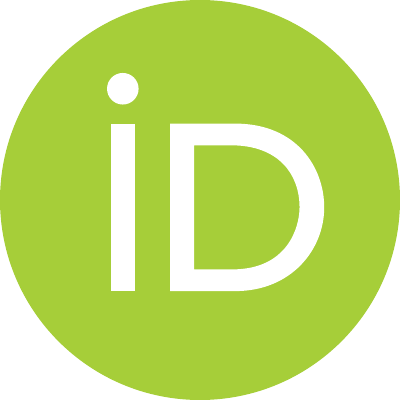}}}} 

\title{\boldmath Observation of the Exotic State $\pi_{1}(1600)$ in $\psi(2S)\rightarrow\gamma\chi_{c1},\chi_{c1}\rightarrow\pi^{+}\pi^{-}\eta'$}

\author{
M.~Ablikim$^{1}$\BESIIIorcid{0000-0002-3935-619X},
M.~N.~Achasov$^{4,c}$\BESIIIorcid{0000-0002-9400-8622},
P.~Adlarson$^{81}$\BESIIIorcid{0000-0001-6280-3851},
X.~C.~Ai$^{87}$\BESIIIorcid{0000-0003-3856-2415},
C.~S.~Akondi$^{31A,31B}$\BESIIIorcid{0000-0001-6303-5217},
R.~Aliberti$^{39}$\BESIIIorcid{0000-0003-3500-4012},
A.~Amoroso$^{80A,80C}$\BESIIIorcid{0000-0002-3095-8610},
Q.~An$^{77,64,\dagger}$,
Y.~H.~An$^{87}$\BESIIIorcid{0009-0008-3419-0849},
Y.~Bai$^{62}$\BESIIIorcid{0000-0001-6593-5665}
% ...
\\
\vspace{0.2cm}
(BESIII Collaboration)\\
\vspace{0.2cm} {\it
$^{1}$ Institute of High Energy Physics, Beijing 100049, People's Republic of China\\
% ...
$^{4}$ Budker Institute of Nuclear Physics SB RAS (BINP), Novosibirsk 630090, Russia\\
% ...
$^{31}$ INFN Sezione di Ferrara, (A)INFN Sezione di Ferrara, I-44122, Ferrara, Italy; (B)University of Ferrara, I-44122, Ferrara, Italy\\
% ...
$^{39}$ Johannes Gutenberg University of Mainz, Johann-Joachim-Becher-Weg 45, D-55099 Mainz, Germany\\
% ...
$^{62}$ Southeast University, Nanjing 211100, People's Republic of China\\
% ...
$^{64}$ State Key Laboratory of Particle Detection and Electronics, Beijing 100049, Hefei 230026, People's Republic of China\\
% ...
$^{77}$ University of Science and Technology of China, Hefei 230026, People's Republic of China\\
% ...
$^{80}$ University of Turin and INFN, (A)University of Turin, I-10125, Turin, Italy; (B)University of Eastern Piedmont, I-15121, Alessandria, Italy; (C)INFN, I-10125, Turin, Italy\\
$^{81}$ Uppsala University, Box 516, SE-75120 Uppsala, Sweden\\
% ...
$^{87}$ Zhengzhou University, Zhengzhou 450001, People's Republic of China\\
\vspace{0.2cm}
$^{\dagger}$ Deceased\\
% ...
$^{c}$ Also at the Novosibirsk State University, Novosibirsk, 630090, Russia\\
% ...
}
}

{
\renewcommand{\baselinestretch}{1.20}

\begin{abstract} 
A partial wave analysis of the process $\psi(2S)\rightarrow\gamma\chi_{c1}, \chi_{c1}\rightarrow\pi^+\pi^-\eta^{\prime}$ is performed using $(2712.4\pm14.3)\times10^{6}$ $\psi(2S)$ events collected with the BESIII detector.
An isovector state with exotic quantum numbers $J^{PC}=1^{-+}$, denoted as $\pi_{1}(1600)$, is observed for the first time 
in the charmonium decay of $\chi_{c1}\rightarrow\pi_{1}^{\pm}(1600)\pi^{\mp}$, $\pi_{1}^{\pm}(1600)\rightarrow\pi^{\pm}\eta^{\prime}$ with a statistical significance over $21\sigma$.
Its mass and width are determined to be
$1828 \pm 8 ({\rm stat})^{+11}_{-33}({\rm syst})~\mathrm{MeV}/c^2$ and $638 \pm 26 ({\rm stat})^{+35}_{-86}({\rm syst})~\mathrm{MeV}$, respectively,  using a relativistic Breit-Wigner function with a mass-dependent width. 
The corresponding product of branching fractions is determined to be
$\mathcal{B}\left[\chi_{c1}\rightarrow\pi_{1}(1600)^{\pm}\pi^{\mp} \right] \times \mathcal{B}\left[\pi_{1}(1600)^{\pm}\rightarrow\pi^{\pm}\eta^{\prime}\right] =  \left( 4.30 \pm 0.14 ({\rm stat})^{+1.04}_{-1.03}({\rm syst})~ \right) \times 10^{-4}$.

\end{abstract}
\vspace{-2.0mm}
}
\maketitle

Within the conventional quark model~\cite{Jaffe:1975fd,Amsler:2004ps,Klempt:2007cp} the hadrons are quark–antiquark bound states (mesons) or three-quark systems (baryons).
The non-Abelian property of quantum chromodynamics (QCD) permits the existence of new types of hadrons, such as glueballs, hybrids, and multiquark states.
Among them, hadrons with quantum numbers that are not allowed in the constituent quark model, such as $J^{PC} = 0^{--}, 0^{+-}, 1^{-+}, 2^{+-}$, etc., 
are manifestly exotic hadrons,  which give essential information for advancing the understanding of QCD.

Lattice QCD (LQCD)~\cite{Lacock:1996ny,MILC:1997usn,Mei:2002ip,Dudek:2013yja} 
and phenomenological models of QCD~\cite{Andreo:1976mf,PhysRevD.17.898,Barnes:1977hg,PhysRevD.86.065013,Burden2002,PhysRevC.55.2649} 
have performed many calculations for the mass spectra of hybrid mesons, which carry valence quark and excited gluonic degrees of freedom.
These models and LQCD predict that the hybrid meson with the lightest mass has the exotic $J^{PC} = 1^{-+}$, and a mass in the range of 1.7 -- 2.1~$\GeV/c^{2}$~\cite{Lacock:1996ny,Dudek:2013yja,review_Meyer:2015eta}.
Although the C parity is not defined for a charged system, it is conventional to quote the $J^{PC}$ quantum numbers of the corresponding neutral partner state in the isospin triplet.
Currently, there are three isovector $1^{-+}$ hybrid candidates around the above mass region, 
namely $\pi_{1}(1400)$, $\pi_{1}(1600)$ and $\pi_{1}(2015)$. 
To account for the two $1^{-+}$ candidates $\pi_{1}(1400)$ and $\pi_{1}(1600)$ in close mass proximity but exhibiting different decay patterns, the coupled-channel analyses~\cite{JPAC_onepole_2019,onepole_2021} suggest a description involving a single resonant pole with separate couplings to the $\pi\eta$ and $\pi\eta^{\prime}$ channels.
The $\pi_{1}(1600)$ is considered one of the most promising candidate, which has been observed in various decay modes, including $\pi\rho$~\cite{PhysRevLett.81.5760, COMPASS_2022}, $\pi f_1(1285)$~\cite{KUHN2004109,Zaitsev:2000rc,Amelin2005}, $\pi b_1(1235)$~\cite{E852_1600_2005,Baker:2003jh,Zaitsev:2000rc,Amelin2005,Khokhlov:2000tk} and $\pi\eta^{\prime}$~\cite{Amelin2005,Khokhlov:2000tk,E852_1600_2001,CLEOc_1600_PhysRevD.84.112009}.
These decay modes have predominantly been studied in diffractive production,
where an artifact may be induced by imperfections in the analysis method~\cite{COMPASS_2022}. 
In contrast, charmonium decays at $e^{+}e^{-}$ colliders provide an ideal environment for searching for hybrid mesons, thanks to well-defined initial states, controllable backgrounds and the ability to fully account for interference through a global unbinned fit.
The $\chi_{c1}\rightarrow\pi\pi\eta^{(\prime)}$ decay is a particularly suitable system for searches for hybrid mesons with $J^{PC} = 1^{-+}$, 
as the $\pi\eta^{(\prime)}$ system in a $S$-wave transition of this decay must exhibit $J^{PC} = 1^{-+}$.
No significant $\pi_{1}(1600)$ signal is observed in the $\chi_{c1}\rightarrow\pi^{+}\pi^{-}\eta$ process by the CLEO~\cite{CLEOc_1600_PhysRevD.84.112009} and BESIII~\cite{besiii_1400_2017} collaborations.
In the $\chi_{c1}\rightarrow\pi^{+}\pi^{-}\eta^{\prime}$ decay using $25.9\times10^{6}$ $\psi(2S)$ events, 
the CLEO collaboration found evidence for an exotic $P$-wave $\pi\eta^{\prime}$ amplitude with a significance of 4$\sigma$, 
but non-resonant $P$-wave $\pi\eta^{\prime}$ interactions could not be excluded~\cite{CLEOc_1600_PhysRevD.84.112009}.
Therefore, further investigation with a larger $\psi(2S)$ data sample is necessary to understand this resonant nature.

In this Letter, we report the first observation and spin-parity determination of the $\pi_{1}(1600)$ in the decay $\psi(2S) \rightarrow \gamma\chi_{c1}, \chi_{c1} \rightarrow \pi^{+}\pi^{-}\eta^{\prime}$
with the two subsequent decay channels of $\eta^{\prime}\rightarrow\gamma\pi^{+}\pi^{-}$ and $\eta^{\prime}\rightarrow\eta\pi^{+}\pi^{-}(\eta\rightarrow\gamma\gamma)$
using $(2712.4\pm14.3)\times 10^{6}$ $\psi(2S)$ events \cite{psip_number} accumulated with the BESIII detector.
A detailed description of the design and performance of the BESIII detector can be found in Ref.~\cite{detector}.

Simulated data samples produced with
{\footnotesize{\sc{GEANT4}}}-based \cite{geant4} Monte
Carlo (MC) software, which includes the geometric description of the
BESIII detector and the detector response, are used to determine
detection efficiencies and to estimate background contributions. The simulation
models the beam energy spread and initial state radiation in the
$e^{+}e^{-}$ annihilations with the generator {\footnotesize{\sc{KKMC}}} \cite{KKMC,KKMC2}.  
An inclusive MC sample includes the production of the
$\psi(2S)$ resonance, the initial state radiation production of the $J/\psi$, and
the continuum processes incorporated in {\footnotesize{\sc{KKMC}}} \cite{KKMC,KKMC2}.
All particle decays are modeled with {\footnotesize{\sc{EVTGEN}}} \cite{EVTGEN,EVTGEN2} using branching fractions either taken from the Particle Data Group (PDG) \cite{pdg}, when available, or otherwise estimated with {\footnotesize{\sc{LUNDCHARM}}} model~\cite{Lund-Charm,Lund-Charm2}.
Final state radiation from charged final state particles is
incorporated using {\footnotesize{\sc{PHOTOS}}} \cite{photos}.
Signal MC samples for the process $\psi(2S) \rightarrow \gamma\chi_{c1}, \chi_{c1} \rightarrow \pi^{+}\pi^{-}\eta^{\prime}$ with the subsequent decays of $\eta^{\prime} \rightarrow \pi^{+}\pi^{-}\eta$ and $\eta \to \gamma\gamma$ are generated uniformly in phase space (PHSP).
A special generator takes $\rho - \omega$ interference and box anomaly into account \cite{etap_generator_2018} in the process of $\eta^{\prime}\rightarrow\gamma\pi^{+}\pi^{-}$.

Charged tracks reconstructed from the multilayer drift chamber (MDC) are required to be within the polar angle range $\vert\!\cos\theta\vert < 0.93$, 
where $\theta$ is defined with respect to the $z$ axis, which is the symmetry axis of the MDC.
The distance of closest approach to the interaction point for charged tracks must be
less than 10 \textrm{cm} along the $z$ axis and less than 1 \textrm{cm} in the transverse plane. 
All charged tracks are assumed to be pions.

Photon candidates are identified using showers in the electromagnetic calorimeter (EMC). 
The deposited energy of each shower is required to have at least 100 MeV both in the barrel region ($\vert\!\cos\theta\vert<0.80$) and the end cap region ($0.86<\vert\!\cos\theta\vert<0.92$). 
To exclude showers from charged tracks, the opening angle between the shower position and the charged tracks extrapolated to the EMC must be greater than $10^{\circ}$. 
The difference between the EMC time and the event start time is required to be within [0, 700] ns in order to suppress electronic noise and energy deposits unrelated to the event.

For the $\psi(2S) \rightarrow \gamma\chi_{c1}, \chi_{c1} \rightarrow \pi^{+}\pi^{-}\eta^{\prime}, \eta^{\prime}\rightarrow\gamma\pi^{+}\pi^{-}$ channel, 
each event candidate is required to have four charged tracks with zero net charge and at least two photons.
A four-constraint (4C) kinematic fit under the $\psi(2S) \rightarrow \gamma\gamma \pi^{+}\pi^{-}\pi^{+}\pi^{-}$ hypothesis is performed by enforcing energy-momentum conservation. 
if more than one combination exist, the one with the smallest $\chi^{2}_{\rm 4C}$ is selected.
The $\chi^{2}_{\rm 4C}$ is required to be less than 40. 
To suppress miscombination, 
an event is retained only if it contains exactly one combination that simultaneously satisfies $|M_{\gamma\pi^{+}\pi^{-}} - m_{\eta^{\prime}}| < 45~\mathrm{MeV}/c^{2}$ and $M_{\gamma\pi^{+}\pi^{-}\pi^{+}\pi^{-}} > 3.4~\mathrm{GeV}/c^{2}$ , where $m_{\eta^{\prime}}$ is the nominal mass of the $\eta^{\prime}$ meson from the PDG~\cite{pdg}.
The $\eta^{\prime}$ and $\chi_{c1}$ candidates are required to satisfy $|M_{\gamma\pi^{+}\pi^{-}} - m_{\eta^{\prime}}| < 15~\mathrm{MeV}/c^{2}$ and $|M_{\gamma\pi^{+}\pi^{-}\pi^{+}\pi^{-}} - m_{\chi_{c1}}| < 15~\mathrm{MeV}/c^{2}$, respectively, where $m_{\chi_{c1}}$ is the $\chi_{c1}$ mass from the PDG~\cite{pdg}.
To suppress non-$\eta^{\prime}$ background, the invariant mass of the $\pi^{+}\pi^{-}$ system (originating from $\eta^{\prime}$) is required to satisfy $M_{\pi^{+}\pi^{-}} >  0.60~\mathrm{GeV}/c^{2}$. 
Events with $|M_{\gamma\gamma} - m_{\pi^{0}}| < 17~\mathrm{MeV}/c^{2}$($|M_{\gamma\pi^{+}\pi^{-}} - m_{\eta}| < 22~\mathrm{MeV}/c^{2}$) are rejected to veto backgrounds containing $\pi^{0}$($\eta$) , where $m_{\pi^{0}}$ and $m_{\eta}$ are the known nominal masses of $\pi^{0}$ and $\eta$ mesons~\cite{pdg}.
To suppress the $J/\psi$ backgrounds from $\psi(2S) \to \gamma\gamma J/\psi \to \gamma \gamma 2(\pi^{+} \pi^{-})$ and $\psi(2S) \rightarrow \pi^{+} \pi^{-} J/\psi$ decays, events are required to satisfy $|M_{2(\pi^{+}\pi^{-})} - m_{J/\psi}| > 28~\mathrm{MeV}/c^{2}$ and $|M_{\pi^{+}\pi^{-}}^{\rm recoil} - m_{J/\psi}| > 6~\mathrm{MeV}/c^{2}$, where $m_{J/\psi}$ is the nominal $J/\psi$ mass from the PDG~\cite{pdg} and $M_{\pi^{+}\pi^{-}}^{\rm recoil}$ is the mass recoiling against the $\pi^{+}\pi^{-}$ system.

For the $\psi(2S) \rightarrow \gamma\chi_{c1}, \chi_{c1} \rightarrow \pi^{+}\pi^{-}\eta^{\prime}, \eta^{\prime}\rightarrow\pi^{+}\pi^{-}\eta,\eta\rightarrow\gamma\gamma$ channel, 
each signal event candidate is required to have four charged tracks with zero net charge and at least three photons.
A 4C kinematic fit is performed under the $J/\psi\rightarrow\gamma\gamma\gamma \pi^{+}\pi^{-}\pi^{+}\pi^{-}$ hypothesis and the combination with the smallest $\chi^{2}_{\rm 4C}$ is chosen.
To suppress miscombination, an event is retained only if it contains exactly one combination that simultaneously satisfies $|M_{\pi^{+}\pi^{-}\gamma\gamma} - m_{\eta^{\prime}}| < 21~\mathrm{MeV}/c^{2}$ and $M_{\gamma\gamma\pi^{+}\pi^{-}\pi^{+}\pi^{-}} > 3.4~\mathrm{GeV}/c^{2}$.
A five-constraint (5C) kinematic fit is performed to further constrain the invariant mass of the two photons to $m_\eta$. 
The resulting $\chi^{2}_{\rm 5C}$ is required to be less than 40. 
The $\eta^{\prime}$ and $\chi_{c1}$ candidates must satisfy $|M_{\pi^{+}\pi^{-}\eta} - m_{\eta^{\prime}}| < 7~\mathrm{MeV}/c^{2}$ and $|M_{\eta\pi^{+}\pi^{-}\pi^{+}\pi^{-}} - m_{\chi_{c1}}| < 15~\mathrm{MeV}/c^{2}$, respectively.

All the above selection criteria are designed to improve signal extraction efficiency and the signal-to-noise ratio.
The final selected data sample contains 24577 and 12952 candidate events for the $\eta^{\prime}\rightarrow\gamma\pi^+\pi^-$ and $\eta^{\prime}\rightarrow\pi^+\pi^-\eta$ channels, respectively.
The selected $\eta^{\prime}$ and $\chi_{c1}$ candidates from the two $\eta^{\prime}$ decay channels are shown in Supplemental Material~\cite{supplemental_material}.

\begin{figure}[tb]
\centering

\vspace{-2.0mm}

\hspace{-3.2mm}
\subfloat{
	\includegraphics[width=0.5\textwidth]{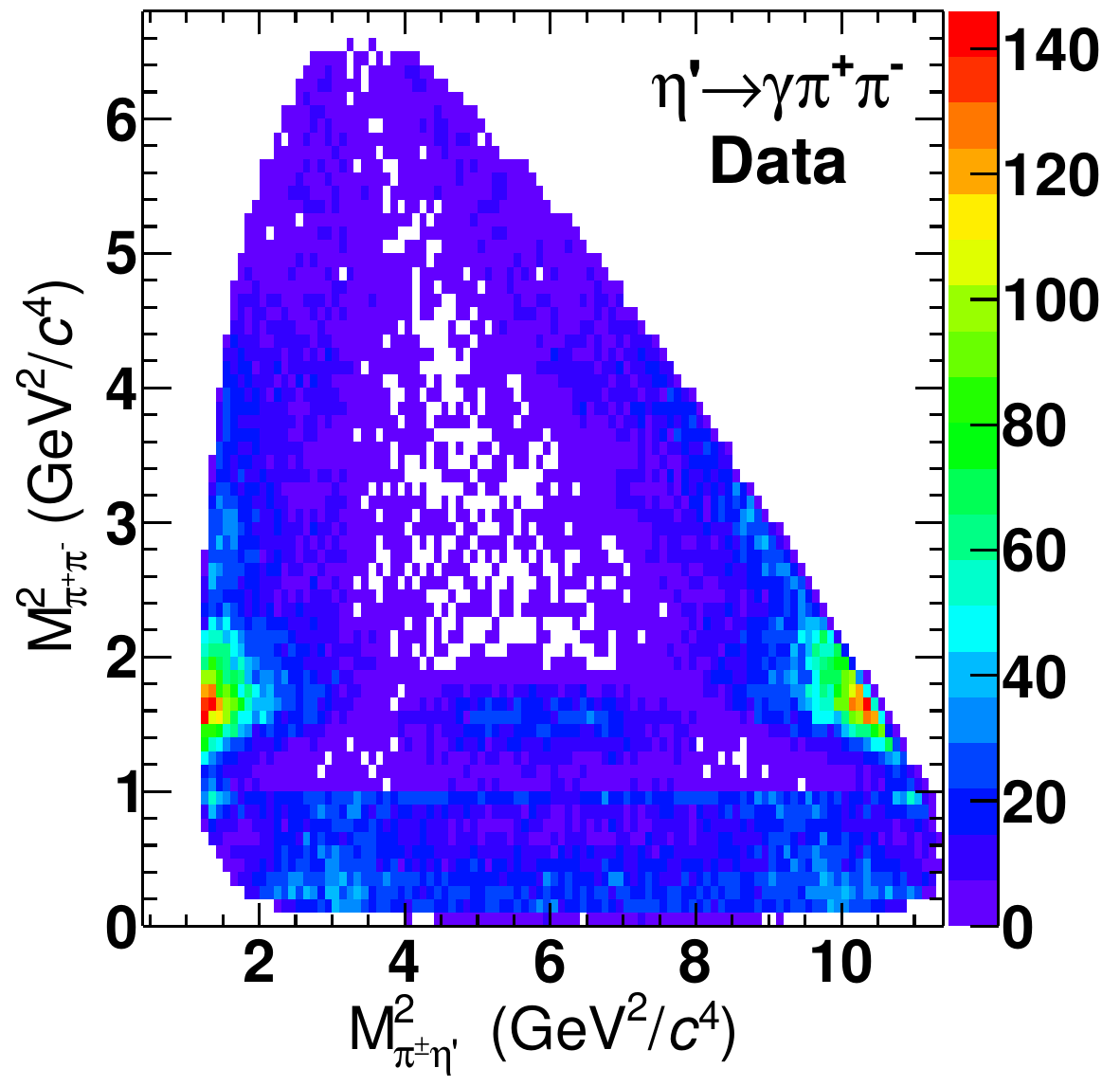}
	\label{fig:Subfigure1}
\put(-107,105){(a)}
}
\hspace{-3.2mm}
\subfloat{
	\includegraphics[width=0.5\textwidth]{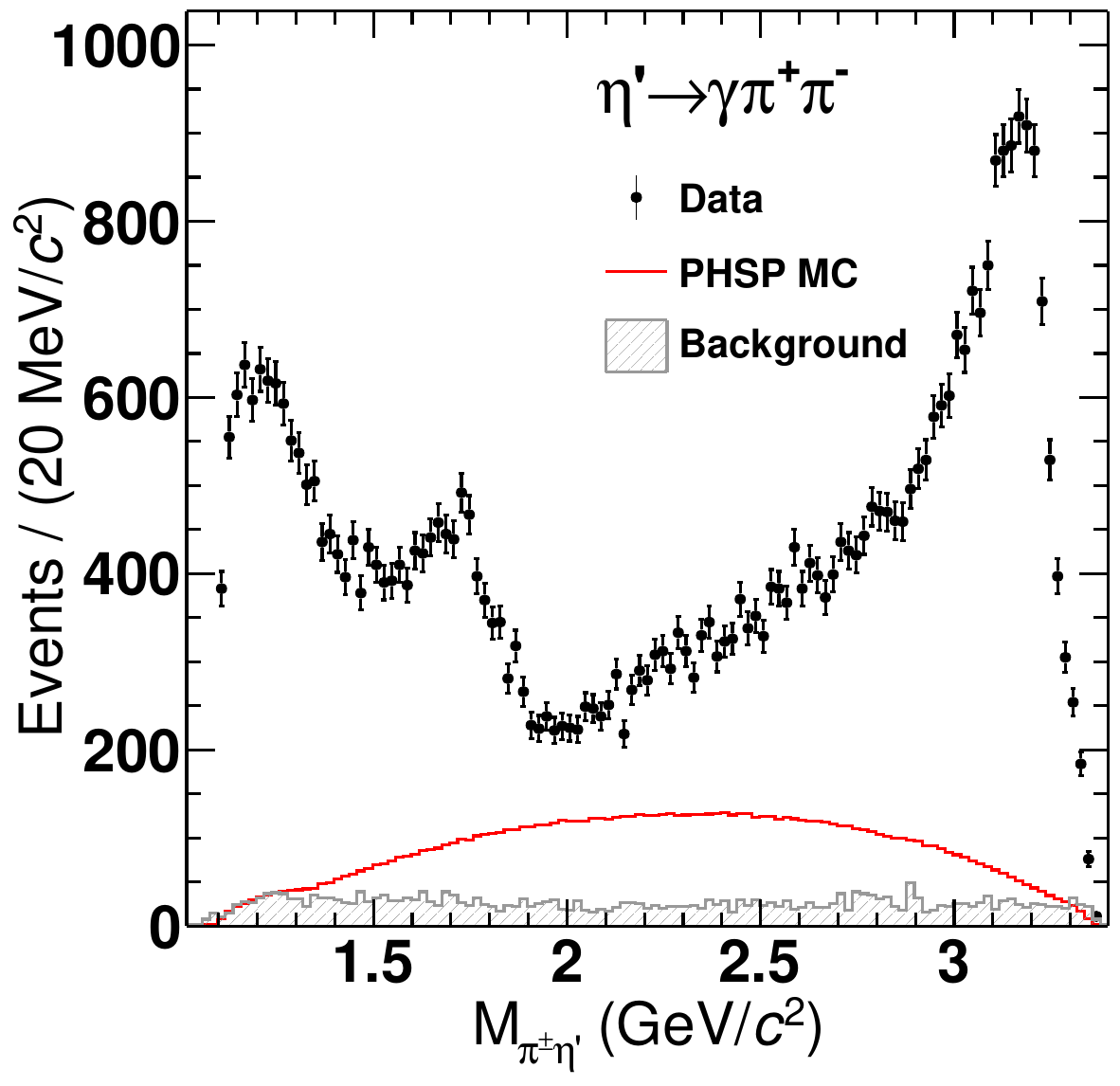}
	\label{fig:Subfigure2}
\put(-102,105){(b)}
}

\vspace{-3.0mm}

\hspace{-3.2mm}
\subfloat{
	\includegraphics[width=0.5\textwidth]{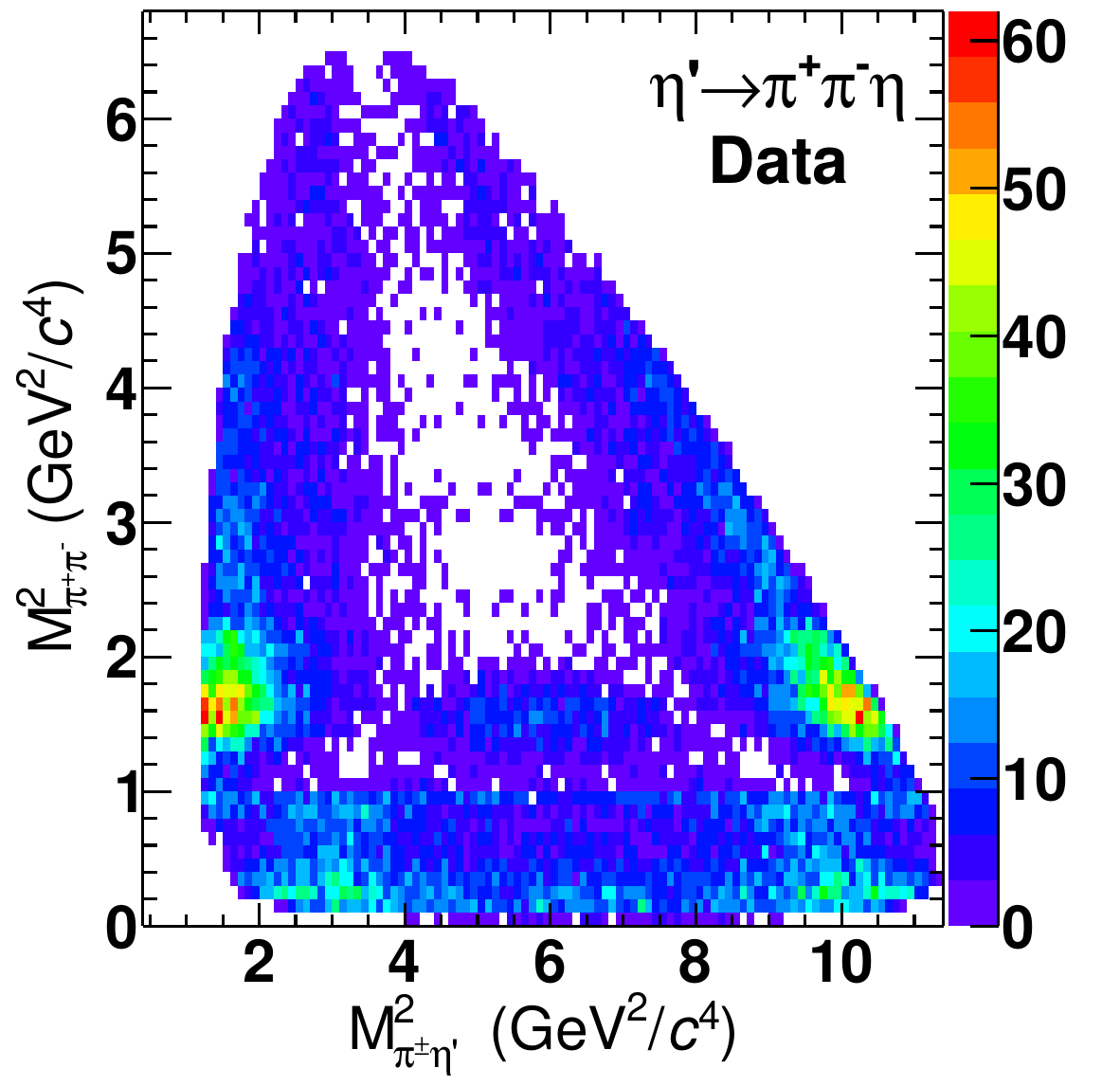}
	\label{fig:Subfigure3}	
\put(-107,105){(c)}
}
\hspace{-3.2mm}
\subfloat{
	\includegraphics[width=0.5\textwidth]{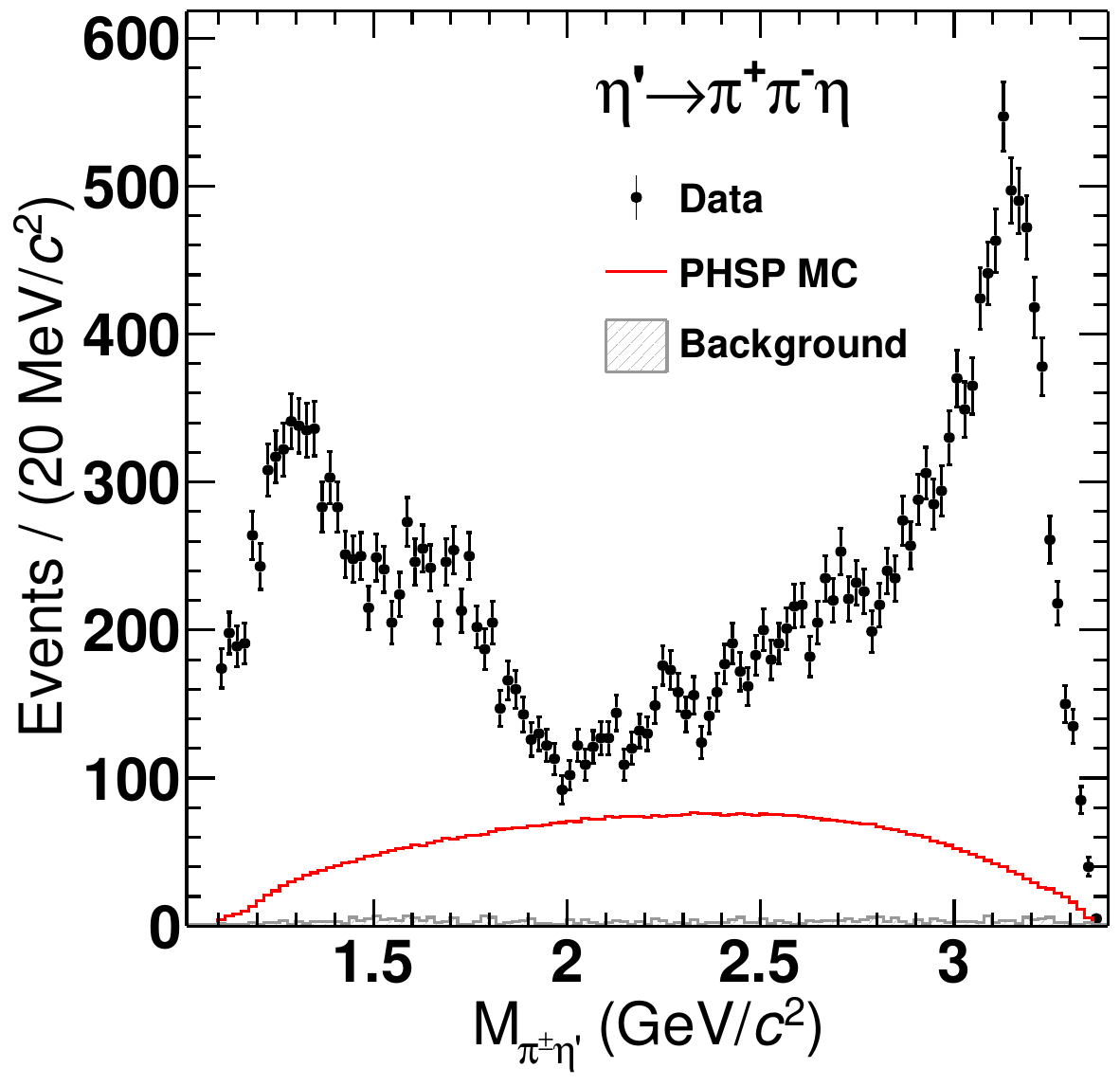}
	\label{fig:Subfigure4}
\put(-102,105){(d)}
}

\caption{
(Left) the Dalitz plot of $M_{\pi^{+}\pi^{-}}^2$ versus $M_{\pi^{\pm}\eta^{\prime}}^2$ 
and (right) the $M_{\pi^{\pm}\eta^{\prime}}$ distributions (two entries for one event)  
for the (top) $\eta^{\prime}\rightarrow\gamma\pi^{+}\pi^{-}$ and  (bottom) $\eta^{\prime}\rightarrow\pi^{+}\pi^{-}\eta$ channels, respectively.
The dots with error bars are data. 
The red histograms are PHSP MC samples with arbitrary normalization. 
The gray shaded histograms are non-$\eta^{\prime}$ background candidates estimated using the $\eta^{\prime}$ sideband.
}

\label{fig:fig1}
\end{figure}

Potential backgrounds are categorized into non-$\eta^{\prime}$ and $\psi(2S) \rightarrow \pi^{0} \pi^{+} \pi^{-} \eta^{\prime}$ processes, both of which have been extensively studied.
The background from non-$\eta^{\prime}$ contribution is estimated using the $\eta^{\prime}$ mass sideband events in the region of
 $30 < |M_{\gamma\pi^{+}\pi^{-}} - m_{\eta^{\prime}}| < 45~\mathrm{MeV}/c^{2}$  
($14 < |M_{\pi^{+}\pi^{-}\eta} - m_{\eta^{\prime}}| < 21~\mathrm{MeV}/c^{2}$) with the corresponding 
fractions of $6.1\%$($1.4\%$) for $\eta^{\prime}\rightarrow\gamma\pi^{+}\pi^{-}$($\eta^{\prime}\rightarrow\pi^{+} \pi^{-} \eta^{\prime}$).
To estimate the background from $\psi(2S) \rightarrow \pi^{0} \pi^{+} \pi^{-} \eta^{\prime}$,
a data sample of this process is selected and reweighted by the ratio of MC efficiencies for the $\psi(2S)\rightarrow\gamma\pi^+\pi^-\eta^\prime$ and $\psi(2S)\rightarrow\pi^0\pi^+\pi^-\eta^\prime$ selections criteria.
All the background estimations are validated in the mass region of $|M_{\pi^{+}\pi^{-}\eta^{\prime}} - m_{\chi_{c0}}| < 15~\mathrm{MeV}/c^{2}$, where the $\chi_{c0} \to \pi^{+}\pi^{-}\eta^{\prime}$ process is strictly forbidden due to CP-parity conservation.
Figures~\ref{fig:Subfigure1} and \ref{fig:Subfigure3} show the Dalitz plots for the selected events in data.  
Figures~\ref{fig:Subfigure2} and \ref{fig:Subfigure4} display the invariant mass spectra of $\pi^{\pm}\eta^{\prime}$.
There are clear structures around $1.7~\mathrm{GeV}/c^{2}$ in the $\pi^{\pm}\eta^{\prime}$ mass spectra for the two $\eta^{\prime}$ decay channels.

A partial wave analysis (PWA) is performed 
using the GPUPWA framework~\cite{gpuframework}
for the selected events from 
$\psi(2S)\rightarrow \gamma \chi_{c1}$, $\chi_{c1}\rightarrow \pi^+\pi^-\eta^{\prime}$ decays.
The signal amplitudes are constructed using the covariant tensor formalism described in Ref.~\cite{zoubs2003} and parameterized as quasi-sequential two-body decays using the isobar model: 
$\psi(2S)\rightarrow \gamma \chi_{c1}$, $\chi_{c1}\rightarrow Y\eta^{\prime}$ or $\chi_{c1}\rightarrow Z \pi^{\mp}$, where $Y$ and $Z$ represent the $\pi^{+}\pi^{-}$ and $\pi^{\pm}\eta^{\prime}$ isobars, respectively. 
Only the $Y$ and $Z$ states with spin $J<3$ are considered.
The relative magnitudes and phases of the signal amplitudes are determined by an unbinned maximum likelihood fit to the combined data of the two $\eta^{\prime}$ decay modes. 
The non-$\eta^{\prime}$ background contribution is taken into account in the fit 
via the subtraction of the negative log-likelihood (NLL) values with the events estimated from the $\eta^{\prime}$ mass sideband region.
The statistical significance of an amplitude is determined from the change in NLL values 
and degrees of freedom of the PWA fit with and without the amplitude.

\begin{figure}[tb]

\centering
\vspace{-3.0mm}

\hspace{-4.0mm}
\subfloat{
  \includegraphics[width=0.5\textwidth]{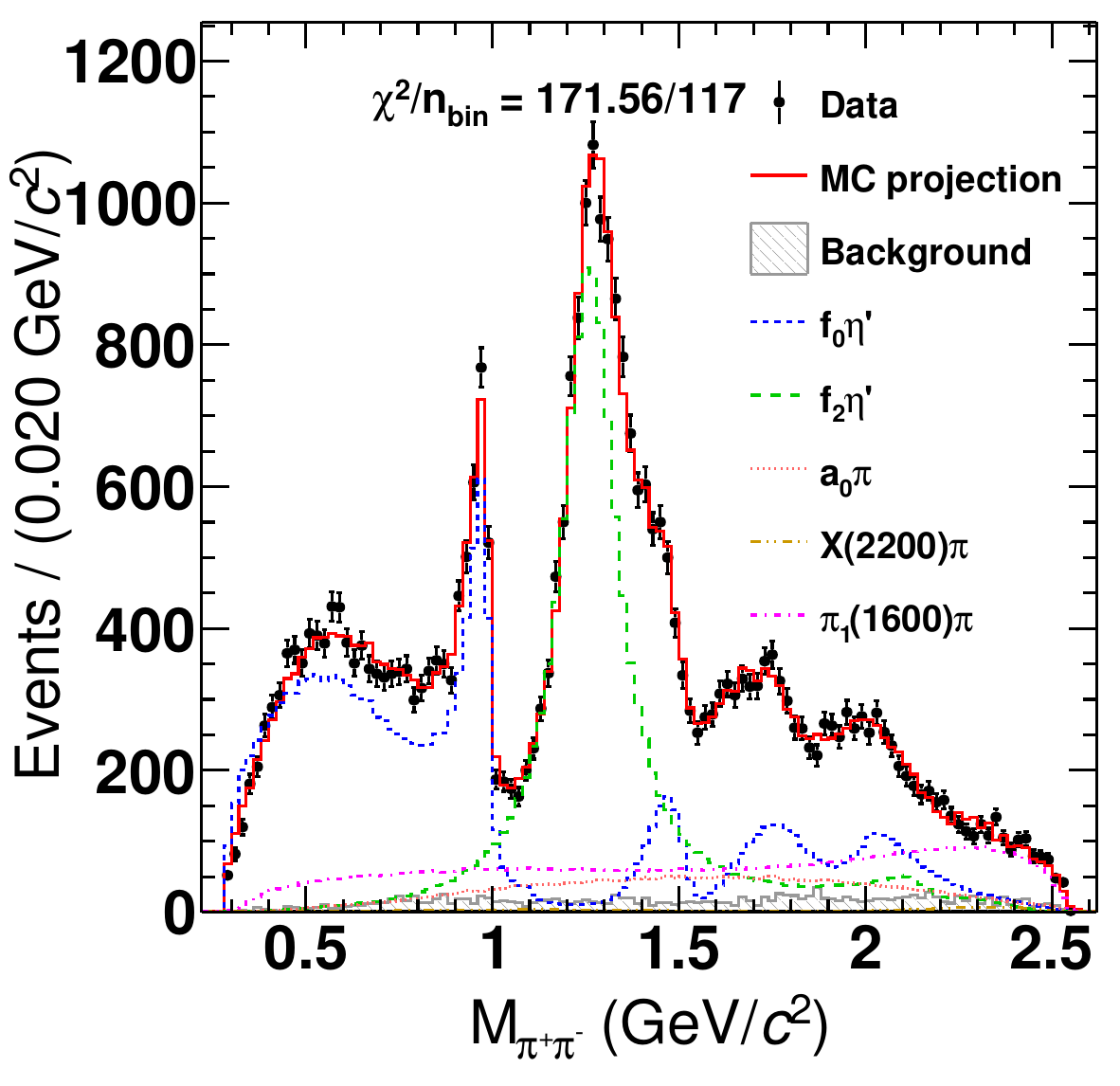}
  \label{fig:Subfigure11}
\put(-100,105){(a)}
}
\hspace{-4.0mm}
\subfloat{
  \includegraphics[width=0.5\textwidth]{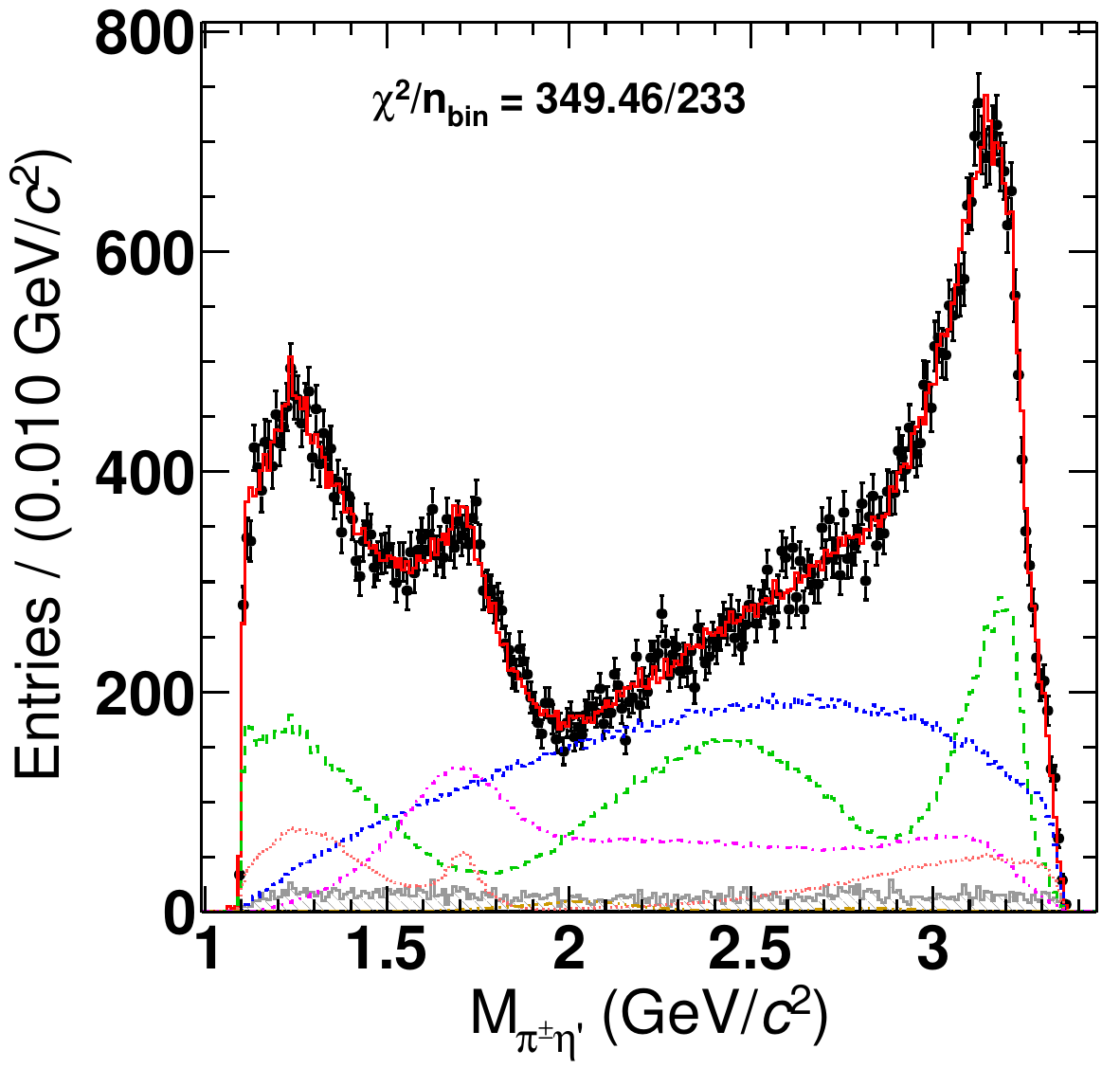}
  \label{fig:Subfigure12}
\put(-100,105){(b)}
}
\vspace{-3.0mm}

\hspace{-4.0mm}
\subfloat{
  \includegraphics[width=0.5\textwidth]{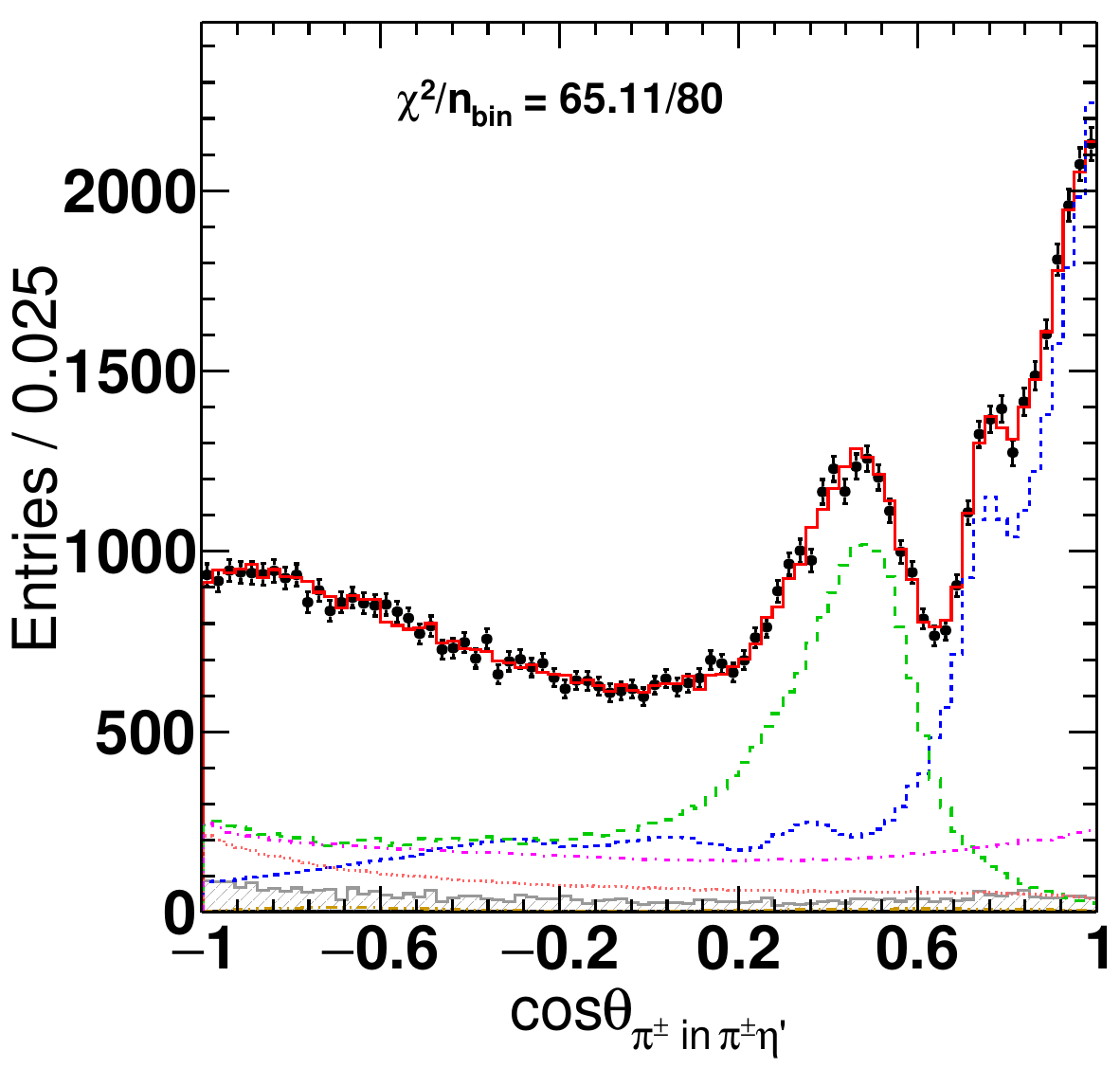}
  \label{fig:Subfigure13} 
\put(-100,105){(c)}
}
\hspace{-4.0mm}
\subfloat{
  \includegraphics[width=0.5\textwidth]{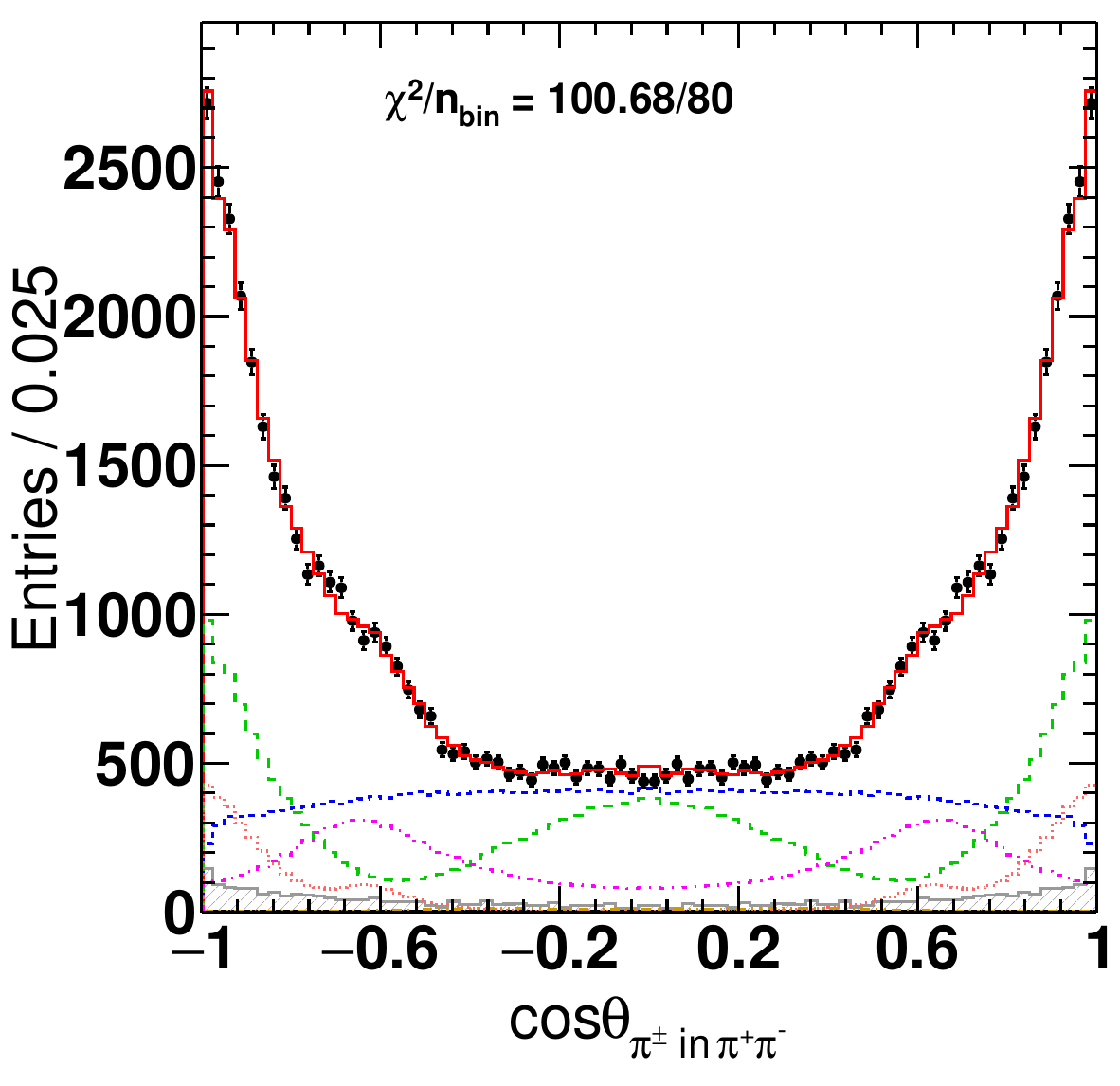}
  \label{fig:Subfigure14}
\put(-100,105){(d)}
}
\vspace{-3.0mm}

 \caption{
  Comparisons between data (with the combination of two $\eta^{\prime}$ decay modes) and PWA fit projections: (a) and (b) are the invariant mass distributions of $\pi^{+}\pi^{-}$ and $\pi^{\pm}\eta^{\prime}$ (two entries for one event), respectively. The angular distributions of $\cos\theta$ (two entries for one event), where $\theta$ is the helicity polar angle of the
  $\pi^{\pm}$ meson in the $\pi^{\pm}\eta^{\prime}$ rest system and the $\pi^{\pm}$ meson in the $\pi^{+}\pi^{-}$ rest system are shown in (c) and (d). 
  The dots with error bars are data. The solid red histograms are the PWA total projections. The shaded gray histograms are the non-$\eta^{\prime}$ background candidates estimated using the $\eta^{\prime}$ sideband. 
  The dashed lines show the coherent sums of $\chi_{c1}$ decay amplitudes corresponding to $\pi^{+}\pi^{-}$ or $\pi^{\pm}\eta^{\prime}$ isobars. }

  \label{fig:pwa}

\end{figure}

The optimal PWA fit can describe the data with a combination of $\chi_{c1}$ decay modes, including: 
$a_0(980)\pi$, $\pi_{1}(1600)\pi$, $a_0(1710)\pi$, $X(2200)\pi$,    
$f_2(1270)\eta^{\prime}$, $f_0(2020)\eta^{\prime}$, $f_2(2150)\eta^{\prime}$, $f_2(2340)\eta^{\prime}$, 
and a component of $(\pi\pi)_{S}\eta^{\prime}$
for the $S$-wave in the $\pi^{+}\pi^{-}$ system. All have statistical significances greater than $5\sigma$.
The fit fractions for each component can be found in Supplemental Material~\cite{supplemental_material}.
The $\pi\pi$ $S$-wave is parameterized using the $K$-matrix formalism\cite{Anisovich:2002ij,BaBar:2008inr}.
More details about the $\pi\pi$ $S$-wave parametrization can be found in Ref.~\cite{besiii_D0_2024}.
The $a_{0}(980)$ line shape is described using dispersion integrals~\cite{besiii_1400_2017}.
The other resonances are described using relativistic Breit-Wigner(BW) functions with mass-dependent widths~\cite{mass_dependent_width}.
For the known resonances contributing less than $3\%$,
the masses and widths of the $a_0(1710)$, $f_2(2150)$, and $f_2(2340)$ are fixed to their PDG values~\cite{pdg},
while the parameters of the $f_0(2020)$ are set to the measurement in Ref.~\cite{besiii_1855_2022}.
The masses and widths of the $f_2(1270)$ and the $\pi_{1}(1600)$ 
are floated in the PWA fit.
The resulting mass and width of $f_2(1270)$ are $1282\pm2\text{(stat)}~\MeV/c^{2}$ and $187\pm4\text{(stat)}~\MeV$, respectively,
which agree with the PDG values~\cite{pdg} within 3 standard deviations.
The mass and width of the $\pi_{1}(1600)$ are determined to be
$\Mpi~\mathrm{MeV}/c^2$ and $\Wpi~\mathrm{MeV}$, respectively.
The corresponding product of branching fractions of the $\pi_{1}(1600)$, including charge conjugate process, is $\mathcal{B}\left[\chi_{c1}\rightarrow\pi_{1}(1600)^{\pm}\pi^{\mp} \right] \times \mathcal{B}\left[\pi_{1}(1600)^{\pm}\rightarrow\pi^{\pm}\eta^{\prime}\right] =  \left( \Bpi~ \right) \times 10^{-4}$
with the statistical significance greater than $21\sigma$.
The BW phase motion is probed by replacing the BW parametrization for the $\pi_{1}(1600)$ with an amplitude equal to the magnitude of the BW function(the phase is constant). 
The significance is found to be greater than $11\sigma$.
A broad $2^{++}$ structure $X(2200)$ in the $\pi^{\pm}\eta^{\prime}$ system is needed with a mass and width of $2206^{+30}_{-32}~\MeV/c^{2}$ and $604^{+169}_{-119}~\MeV$, respectively, with a statistical significance of $8\sigma$ and a fractional contribution of approximately $0.8\%$.
The $X(2200)$ may be a small combined effect of several $2^{++}$ states.
The impact of excluding the $X(2200)$ is treated as a source of systematic uncertainty.
Figure~\ref{fig:pwa} shows the comparison of the mass and angular distributions between the data and the optimal PWA fit, 
demonstrating a good agreement with the data.
The $\chi^2\mathrm{/n_{bin}}$ value is displayed on each figure to demonstrate the fit quality.

The PWA fit is tested against all possible additional PDG states, unknown resonances, and non-resonant contributions.
All PDG-listed states are found to have a significance below 3$\sigma$, except for the $a_2(1320)$, which has a significance of 3.9$\sigma$.
Although this state is not included in the nominal solution, its potential contribution is accounted for as a source of systematic uncertainty.
The scan results yield no evidence for extra intermediate states.
For the spin-parity determination of the $\pi_{1}(1600)$, the $1^{-+}$ hypothesis yields a better fit than the $0^{++}$ or $2^{++}$ hypotheses, with significances exceeding $17\sigma$.
The significance of the $1^{-+}$ over other alternative $J^{PC}$ assumptions is 
determined from the changes in the NLL value and the number of degrees of freedom of the PWA fit and 
evaluated with the consideration of all systematic uncertainty variations as described below.
The quantum numbers of the other resonances in the optimal PWA fit are consistent with those listed in the PDG~\cite{pdg}.

Systematic uncertainties associated with the event selection, 
including tracking~\cite{tracking_uncertainty}, photon selection~\cite{PhysRevD.81.052005}, 
kinematic fit~\cite{helix_method},
the $\eta^{\prime}$ and $\chi_{c1}$ mass window requirements,
the branching fractions of $\psi(2S)\rightarrow\gamma\chi_{c1}$~\cite{pdg}, 
 $\eta^{\prime}\rightarrow\pi^{+}\pi^{-}\eta$ ($\eta\rightarrow\gamma\gamma$)  and $\eta^{\prime}\rightarrow\gamma\pi^{+}\pi^{-}$~\cite{pdg}
and the total number of $\psi(2S)$ events \cite{psip_number}, only affect the branching fraction measurement of the $\pi_{1}(1600)$.
Uncertainties due to these effects are estimated to be $\pm6\%$.

Systematic uncertainties associated with the PWA affect measurements of the branching fraction and the resonance parameters of the $\pi_{1}(1600)$. 
Sources of uncertainty include 
the non-$\eta^{\prime}$ background estimation, 
$\chi_{c2}\rightarrow\pi^+\pi^-\eta^{\prime}$ processes contribution,
resonance parametrization,
resonance combination.
The uncertainty from the background estimation is determined by varying the non-$\eta^{\prime}$ background normalization factors by 30$\%$.
The possible impact from the $\chi_{c2}$ meson contribution is taken into account by including the $\chi_{c2} \to \pi^+\pi^-\eta^{\prime}$ processes in the PWA, 
even though it has a negligible statistical significance and small fractional contribution.
The uncertainties from the resonance parameterization are determined by varying the resonance parameters and line shapes.
Systematic uncertainties due to the $f_{2}(1270)$, $a_{0}(1710)$, $f_{0}(2020)$ and $a_{0}(980)$ mass and width parameters are evaluated by varying each within one standard deviation of the reference values~\cite{pdg,besiii_1855_2022,besiii_1400_2017} and floating these parameters in the fit.
To estimate the uncertainty from $\pi\pi$ $S$-wave parametrization, 
the contribution of $K$-matrix is replaced by a combination of multi-resonance states, including $f_{0}(500)$, $f_{0}(980)$, $f_{0}(1370)$, $f_{0}(1500)$, $f_{0}(1710)$ mesons, and a non-resonant $S$-wave in the $\pi^{+}\pi^{-}$ system.
The $f_{0}(500)$ and $f_{0}(980)$ mesons are parameterized using the BESII measurements~\cite{f0_500,980resonances2005}, while the others are parameterized by a relativistic BW function with a mass-dependent width~\cite{mass_dependent_width}.
The resonance combination uncertainty includes the impact from the $a_2(1320)$, $X(2200)$ and $f_{2}$ components($f_{2}(2150)$ and $f_{2}(2340)$).
The systematic uncertainty associated with the $a_2(1320)$ is estimated by including this state in the PWA fit.
The uncertainty related to the $X(2200)$ is evaluated by excluding this state from the fit, due to its small fractional contribution. 
To account for the uncertainty from the description of the $f_2$ components,
the combination of the $f_{2}(2150)$ and $f_{2}(2340)$ is replaced by two of the known $f_2$ mesons among $f_{2}(1950)$, $f_{2}(2150)$, $f_{2}(2300)$ and $f_{2}(2340)$,  due to large overlap between the broad $f_2$ mesons listed in the PDG~\cite{pdg} within this mass region.
The $f_2$ components of the optimal PWA fit is also tested with replacement of a broad $f_2$ resonance.
In addition to the replacements, variations of the masses and widths of the $f_{2}(2150)$ and $f_{2}(2340)$ within one standard deviation of the values reported in the PDG~\cite{pdg} are also included.
The largest deviation from the nominal PWA solution is assigned as the systematic uncertainty arising from the combination of the $f_{2}$ components.

All sources of systematic uncertainty and their contributions are summarized in Table~\ref{tab:sys} and are treated independently. 
Finally, using a BW function with a mass-dependent width,
the mass and width of $\pi_{1}(1600)$ are determined to be $\Mpi\Mpisys~\mathrm{MeV}/c^2$ and $\Wpi\Wpisys~\mathrm{MeV}$, respectively. 
The corresponding product of branching fractions is $\mathcal{B}\left[\chi_{c1}\rightarrow\pi_{1}(1600)^{\pm}\pi^{\mp} \right] \times \mathcal{B}\left[\pi_{1}(1600)^{\pm}\rightarrow\pi^{\pm}\eta^{\prime}\right]$ 
= $\left( \Bpi\Bpisys~ \right) \times 10^{-4}$.
By requiring the denominator in the BW propagator of the $\pi_{1}(1600)$ to be zero,
its pole position is determined to be 
$\left( 1690\pm16^{+36}_{-44} \right) - i\left( 217\pm5^{+7}_{-19} \right) \MeV$,
where the first and second uncertainties are statistical and systematic, respectively.

An alternative analysis using the BW function with a constant width for the $\pi_{1}(1600)$ is re-performed with the consideration of all systematic uncertainties.
This parameterization yields a mass of $\MpiC\MpisysC~\mathrm{MeV}/c^2$ and a width of $\WpiC\WpisysC~\mathrm{MeV}$ for the $\pi_{1}(1600)$.
The product of branching fractions $\mathcal{B}\left[\chi_{c1}\rightarrow\pi_{1}(1600)^{\pm}\pi^{\mp} \right] \times \mathcal{B}\left[\pi_{1}(1600)^{\pm}\rightarrow\pi^{\pm}\eta^{\prime}\right] $ 
is measured to be $\left( \BpiC\BpisysC~ \right) \times 10^{-4}$, with a statistical significance exceeding $21\sigma$. The significance of the BW phase motion is $11\sigma$, and the $1^{-+}$ hypothesis is favorable over the $0^{++}$ or $2^{++}$ hypotheses with significances greater than $17\sigma$.
The corresponding pole position is determined to be
$\left( 1689\pm10^{+33}_{-35} \right) - i\left( 220\pm7^{+9}_{-27} \right) \MeV$, 
where the first uncertainties are statistical and the second are systematic.
Although the BW mass and width of the $\pi_{1}(1600)$ differ between the nominal and alternative parameterizations, its line shape, fractional contribution, phase-motion significance, and $J^{PC}$ assignment remain practically unchanged.
Both parametrizations yield consistent pole positions, indicating that the two BW models for the $\pi_{1}(1600)$ are physically equivalent.
The measured pole mass is also consistent with the results from CLEO~\cite{CLEOc_1600_PhysRevD.84.112009}, JPAC~\cite{JPAC_onepole_2019}, and COMPASS~\cite{COMPASS_2018} measurements.

\begin{table}[tb]
\centering
\renewcommand\arraystretch{1.2}
\caption{Systematic uncertainties on the measurements of mass, width and product branching fraction of the $\pi_{1}(1600)$.}

\resizebox{\columnwidth}{!}{

    \begin{tabular}{l|ccc}
    \hline\hline
    Sources & $\Delta M~(\mathrm{MeV}/c^{2})$ & $\Delta \Gamma~(\mathrm{MeV})$ & $\Delta \mathcal{B}/\mathcal{B}\ (\%)$ \\
    \hline

    Event selection                 & $\cdot\cdot\cdot$     & $\cdot\cdot\cdot$  & $\pm 6$           \\
    Non-$\eta^{\prime}$ background  & $-2$                  & $^{+12}_{-18}$     & $^{+4}_{-5}$      \\
    $\chi_{c2}$ contribution        & $+2$                  & $+5$               & $+2$              \\
    Resonance parametrization       & $^{+5}_{-21}$         & $^{+25}_{-58}$     & $^{+13}_{-22}$    \\
    Resonance combination           & $^{+10}_{-26}$        & $^{+21}_{-61}$     & $^{+19}_{-5}$     \\
    \hline 
    Total   & \MpisysR   & \WpisysR    & \BpisysRP \\
    
    \hline\hline
    \end{tabular}
   }

\label{tab:sys}
\end{table}

In summary, a PWA of $\psi(2S)\rightarrow\gamma\chi_{c1},\chi_{c1}\rightarrow\pi^+\pi^-\eta^{\prime}$ has been performed based on $(2712.4\pm14.3)\times 10^{6}$ $\psi(2S)$ events \cite{psip_number} collected with the BESIII detector.
An isovector exotic state, the $\pi_{1}(1600)$, is observed for the first time in charmonium decays, in the $\pi^{\pm}\eta^{\prime}$ system,
with a statistical significance greater than $21\sigma$. 
Its spin-parity quantum numbers are determined to be $1^{-+}$ with a significance greater than $17\sigma$.
The significance of the BW phase motion is greater than $11\sigma$.
Two BW parametrizations, employing a mass-dependent width and a constant width, respectively, are used to describe the $\pi_{1}(1600)$ 
and they yield a consistent pole position.
The corresponding measurements of the mass, width, and product of branching fractions are provided for each parametrization.
The measured mass of $\pi_{1}(1600)$ is also compatible with LQCD calculations for the predicted $1^{-+}$ hybrid with its mass around $1.7~\GeV/c^{2}$ ~\cite{review_Meyer:2015eta}.
Recently, an isoscalar $1^{-+}$ state, the $\eta_{1}(1855)$, has been observed in the $J/\psi\rightarrow\gamma\eta\eta^{\prime}$ decay channel~\cite{besiii_1855_2022}.
To clarify the nature of the $\pi_{1}(1600)$ and determine whether the $\eta_{1}(1855)$ and $\pi_{1}(1600)$ can be classified into a single $1^{-+}$ hybrid nonet, further investigation of their production mechanisms and decay modes in additional channels is required.
Additionally, the search for the neutral $\pi_{1}(1600)$ in the $\psi(2S)\rightarrow\gamma\chi_{c1},\chi_{c1}\rightarrow\pi^{0}\pi^{0}\eta^{\prime}$ process is important for testing isospin symmetry and confirming its C parity in charmonium decays.

The BESIII Collaboration thanks the staff of BEPCII (https://cstr.cn/31109.02.BEPC) and the IHEP computing center for their strong support. This work is supported in part by National Key R\&D Program of China under Contracts Nos. 2025YFA1613900, 2023YFA1606000, 2023YFA1606704; National Natural Science Foundation of China (NSFC) under Contracts Nos. 11635010, 11935015, 11935016, 11935018, 12025502, 12035009, 12035013, 12061131003, 12192260, 12192261, 12192262, 12192263, 12192264, 12192265, 12221005, 12225509, 12235017, 12342502, 12361141819; the Chinese Academy of Sciences (CAS) Large-Scale Scientific Facility Program; the Strategic Priority Research Program of Chinese Academy of Sciences under Contract No. XDA0480600; CAS under Contract No. YSBR-101; 100 Talents Program of CAS; The Institute of Nuclear and Particle Physics (INPAC) and Shanghai Key Laboratory for Particle Physics and Cosmology; ERC under Contract No. 758462; German Research Foundation DFG under Contract No. FOR5327; Istituto Nazionale di Fisica Nucleare, Italy; Knut and Alice Wallenberg Foundation under Contracts Nos. 2021.0174, 2021.0299, 2023.0315; Ministry of Development of Turkey under Contract No. DPT2006K-120470; National Research Foundation of Korea under Contract No. NRF-2022R1A2C1092335; National Science and Technology fund of Mongolia; Polish National Science Centre under Contract No. 2024/53/B/ST2/00975; STFC (United Kingdom); Swedish Research Council under Contract No. 2019.04595; U. S. Department of Energy under Contract No. DE-FG02-05ER41374.

\vspace{-0.2cm}

\bibliographystyle{apsrev4-2}
\bibliography{draft}

\newpage
\onecolumngrid
\begin{center}
\section*{\boldmath Supplemental Material}
\end{center}

\begin{figure}[H]
\centering

\hspace{-3.2mm}
\subfloat{
	\includegraphics[width=0.25\textwidth]{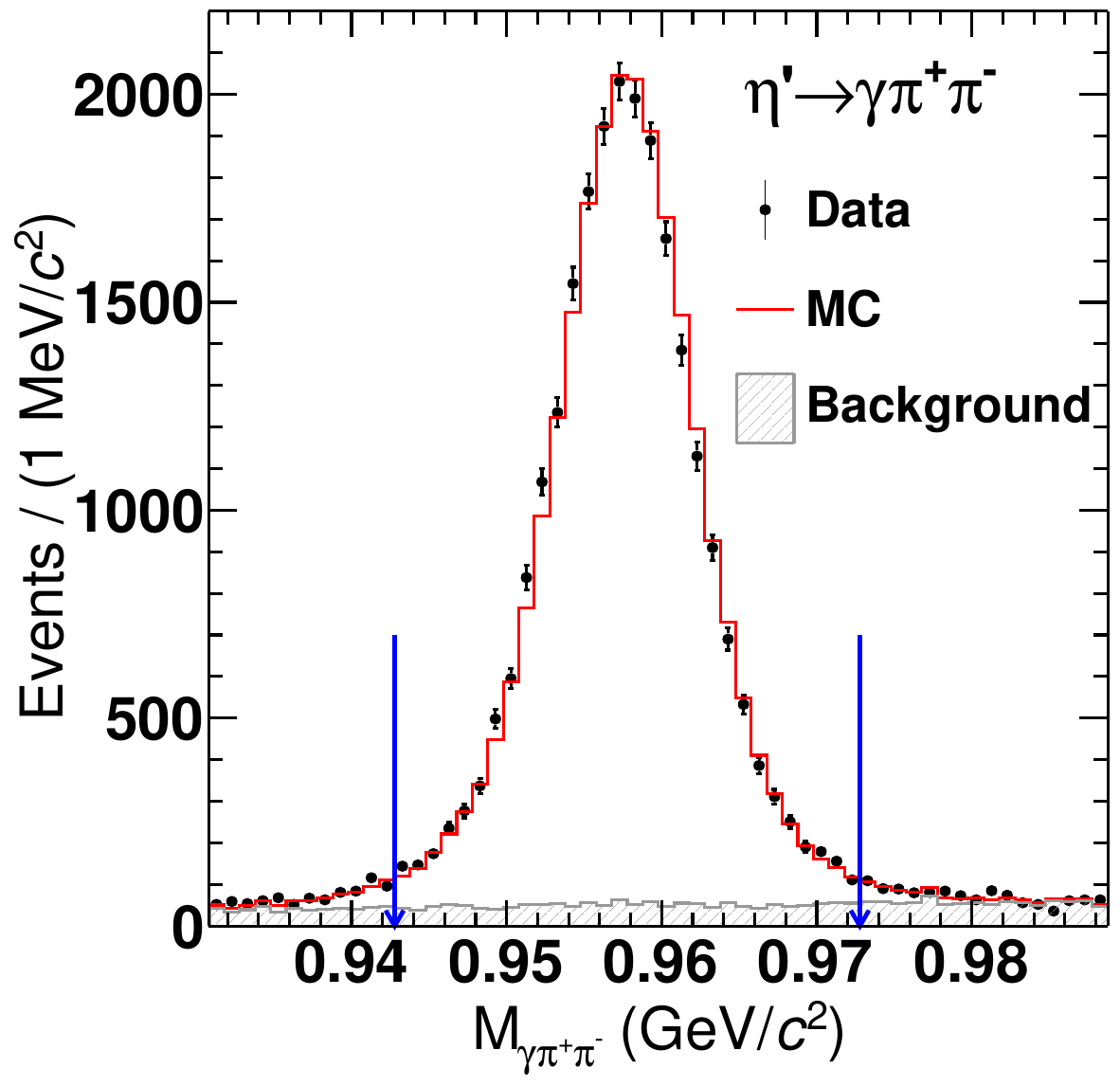}
	\label{fig:Subfigure3a}
\put(-100,105){(a)}
}
\hspace{-3.2mm}
\subfloat{
	\includegraphics[width=0.25\textwidth]{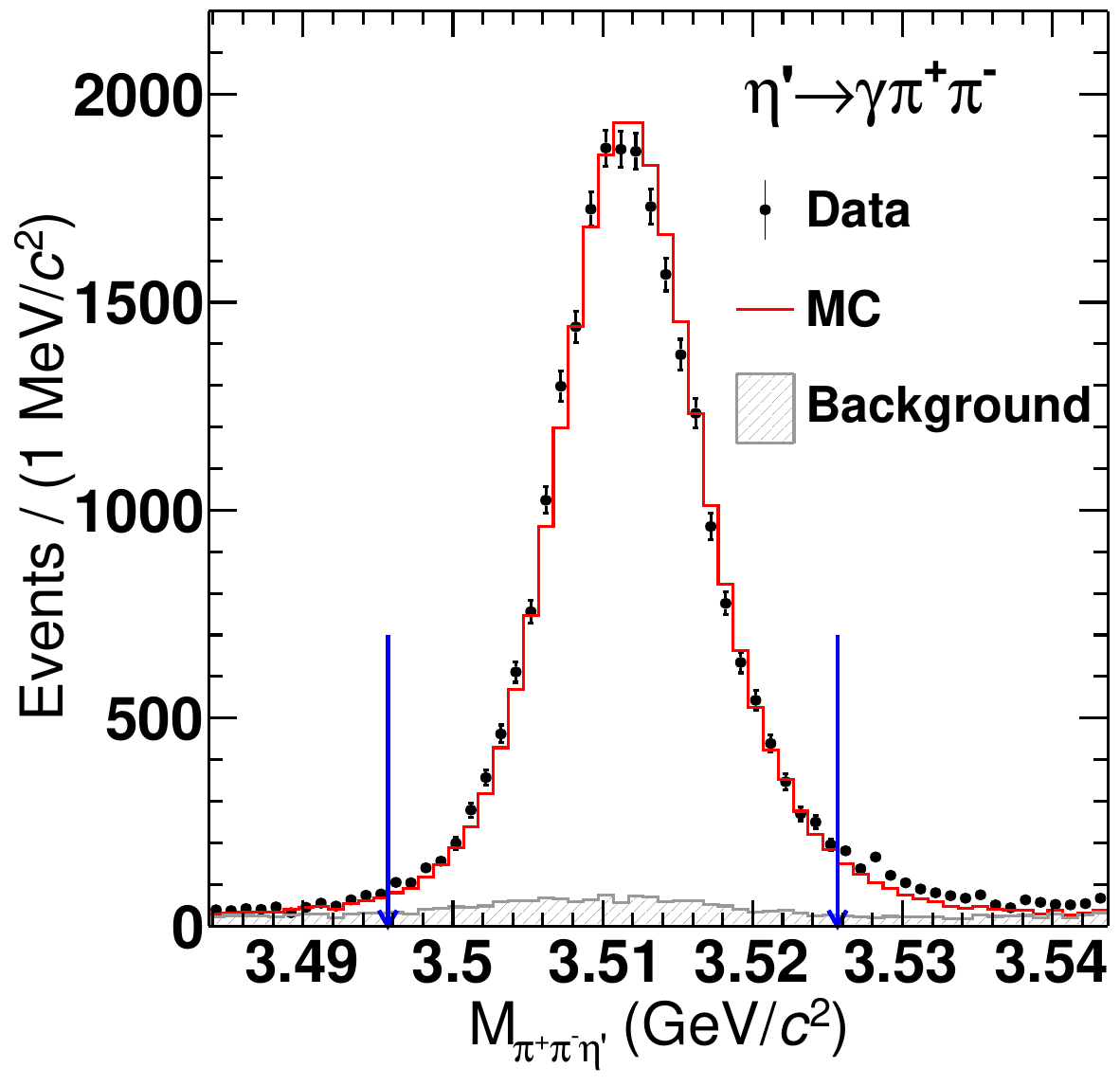}
	\label{fig:Subfigure3b}
\put(-100,105){(b)}
}\\
\hspace{-3.2mm}
\subfloat{
	\includegraphics[width=0.25\textwidth]{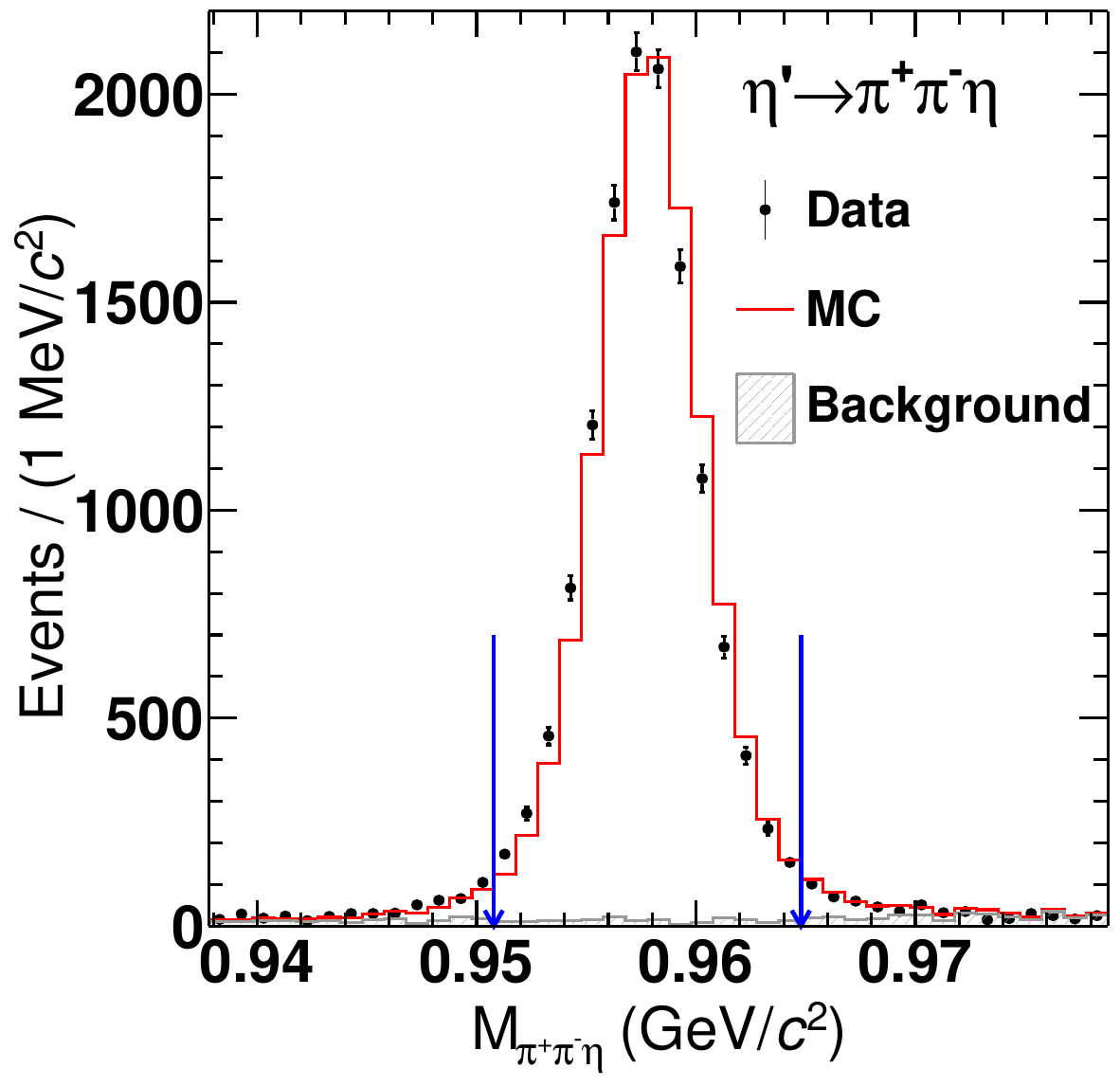}
	\label{fig:Subfigure3c}	
\put(-100,105){(c)}
}
\hspace{-3.2mm}
\subfloat{
	\includegraphics[width=0.25\textwidth]{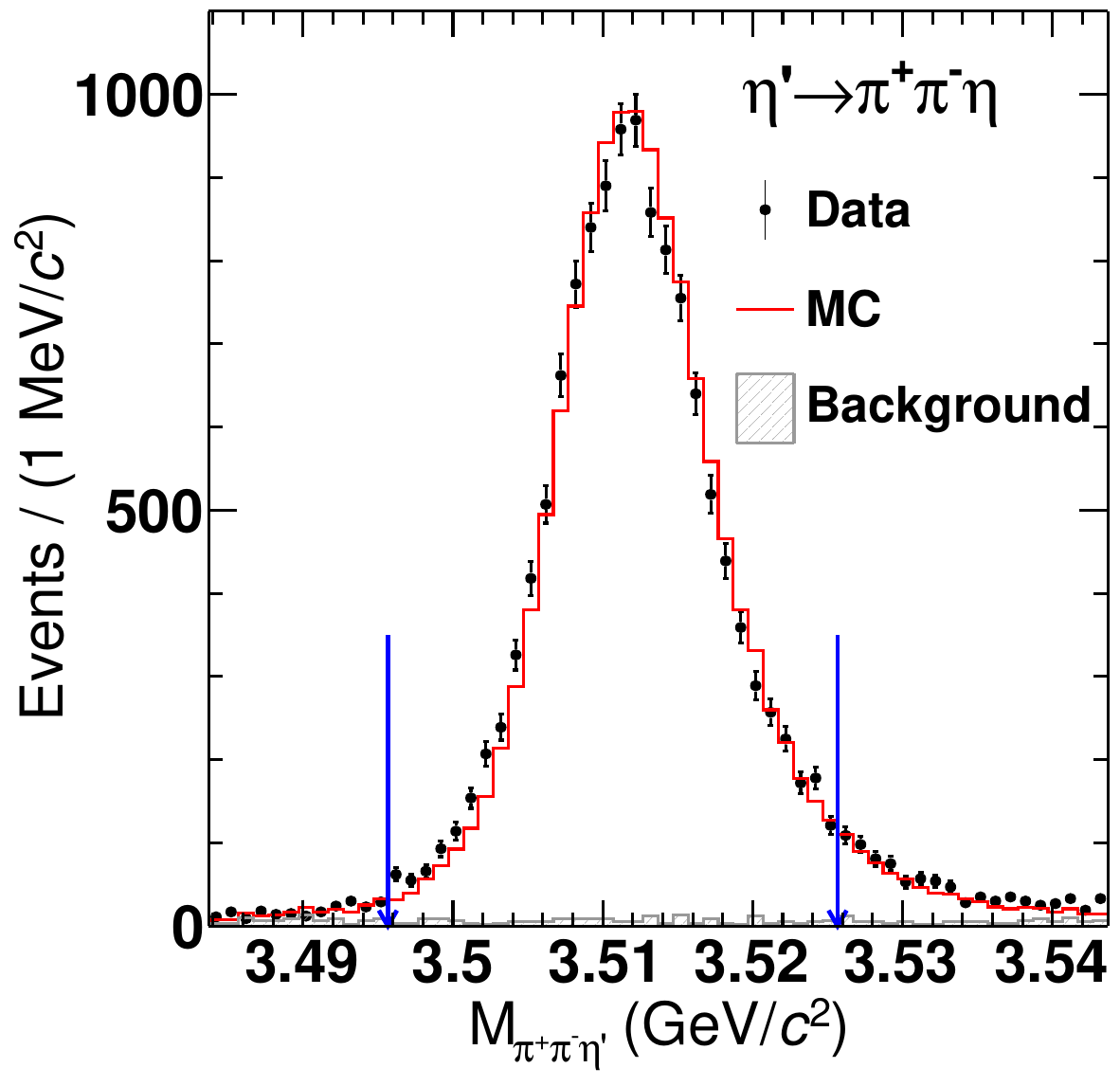}
	\label{fig:Subfigure3d}
\put(-100,105){(d)}
}

\caption{
The invariant mass distributions of (left) $\eta^\prime$ and (right) $\chi_{c1}$ candidates for the (top) $\eta^\prime\to \gamma\pi^+\pi^-$ and (bottom) $\eta^\prime \to \pi^+\pi^-\eta$ channels, respectively.
The dots with error bars are data. 
The shaded gray histograms show the non-$\eta^{\prime}$ background described with the inclusive MC sample, 
after normalizing to the yields estimated in the data.
The red histograms are the sum of PHSP MC and non-$\eta^{\prime}$ background normalized to data yields.
The events between the blue arrows are the selected  $\eta^{\prime}$ and $\chi_{c1}$ signal events.
}

\label{fig:fig3}
\end{figure}

Figs.~\Cref{fig:Subfigure3a,fig:Subfigure3b,fig:Subfigure3c,fig:Subfigure3d} shows the selected $\eta^{\prime}$ and $\chi_{c1}$ candidates from the two $\eta^{\prime}$ decay channels.

The fit fractions for each component and their interference fractions are shown in Table.\ref{tab:pwa_inte}. 

\begin{table}[htb]
	\centering
	\renewcommand\arraystretch{1.2}
	\caption{The fit fractions for each component and the interference fractions between two components($\%$) in the PWA fit. The uncertainties are statistical only.}

		\begin{tabular}{c|c|c|c|c|c|c|c|c|c}
			\hline\hline
			component(\%) & $(\pi\pi)_{S-\text{wave}}$ & $a_{0}(980)$ & $f_{2}(1270)$  & $\pi_{1}(1600)$  & $f_{0}(2020)$ &  $a_{0}(1710)$ & $f_{2}(2340)$ & $f_{2}(2150)$  & $X(2200)$\\
			\hline
$(\pi\pi)_{S-\text{wave}}$ & 40.4$\pm$0.6  & 0.3$\pm$0.3  & -1.5$\pm$0.1  & -6.8$\pm$0.5  & -0.4$\pm$0.3  & 1.1$\pm$0.1  & -0.1$\pm$0.0  & -0.1$\pm$0.0  & -0.3$\pm$0.2 \\
\hline
$a_{0}(980)$ &  --  & 6.6$\pm$0.3  & 5.9$\pm$0.2  & -0.7$\pm$0.1  & 2.1$\pm$0.1  & 1.7$\pm$0.1 & 2.5$\pm$0.2  & -0.0$\pm$0.1  & 0.0$\pm$0.1 \\
\hline
$f_{2}(1270)$ &  --  &  --  & 32.6$\pm$0.5  & -7.8$\pm$0.3  & -0.3$\pm$0.0  & 1.7$\pm$0.1  & -0.8$\pm$0.3  & -0.5$\pm$0.2  & 0.5$\pm$0.1 \\
\hline
$\pi_{1}(1600)$ &  --  &  --  &  --  & 19.3$\pm$0.6  & -0.8$\pm$0.3  & -1.5$\pm$0.1  & -1.0$\pm$0.2  & 0.9$\pm$0.1  & 0.2$\pm$0.1 \\
\hline
$f_{0}(2020)$ &  --  &  --  &  --  &  --  & 1.9$\pm$0.2  & 0.3$\pm$0.0  & -0.1$\pm$0.0  & -0.0$\pm$0.0  & -0.3$\pm$0.1 \\
\hline
$a_{0}(1710)$ &  --  &  --  &  --  &  --  &  --  & 1.3$\pm$0.1  & -0.2$\pm$0.1  & -0.2$\pm$0.0  & -0.0$\pm$0.0 \\
\hline
$f_{2}(2340)$ &  --  &  --  &  --  &  --  &  --  &  --  & 2.2$\pm$0.4  & 0.5$\pm$0.2  & 0.3$\pm$0.1 \\
\hline
$f_{2}(2150)$ &  --  &  --  &  --  &  --  &  --  &  --  &  --  & 0.6$\pm$0.1  & -0.2$\pm$0.1 \\
\hline
$X(2200)$ &  --  &  --  &  --  &  --  &  --  &  --  &  --  &  --  & 0.8$\pm$0.1 \\

			\hline\hline
		\end{tabular}

	\label{tab:pwa_inte}
\end{table}

\twocolumngrid

\end{document}